\documentclass[12pt,letterpaper]{article}

\usepackage[includeheadfoot,
            marginratio={1:1,2:3}, 
            width=412pt,
            height=688pt,]{geometry}

\usepackage[utf8]{inputenc}
\usepackage{amsmath}
\usepackage{amsfonts}
\usepackage{amssymb}
\usepackage{graphicx}
\usepackage{booktabs}
\usepackage{empheq}
\usepackage{paralist}
\usepackage{cite}
\usepackage[hidelinks]{hyperref}
\usepackage{xcolor}
\usepackage{tensor}
\usepackage{tikz}
\usepackage{adjustbox}
\usepackage{subcaption}
\usepackage{braket}
\usetikzlibrary{decorations.pathreplacing,calc}
\usepackage{dsfont}
\usepackage[cal=cm]{mathalfa}
\usepackage{hhline}
\usepackage{tabu}
\usepackage{multirow}

\newcommand{\nc}{\newcommand}
\nc{\lb}{\llbracket}
\nc{\rb}{\rrbracket}
\nc{\gl}{\llbracket}
\nc{\gr}{\rrbracket}
\nc{\del}{\partial}
\nc{\tri}{\hspace{-3.5pt}\vartriangle\hspace{-3.5pt}}
\nc{\blacktri}{\blacktriangle}
\nc{\eq}[1]{\begin{equation}
                     \begin{split} #1 \end{split}
                     \end{equation}}
\nc{\ul}{\underline}
\nc{\ov}{\overline}
\nc{\fa}{\hat}
\nc{\fb}{\MakeUppercase}
\nc{\fc}{\tilde }
\nc{\Lie}{{\cal L}} 
\nc{\lambdabar}{{\mkern0.75mu\mathchar '26\mkern -9.75mu\lambda}}

\newmuskip\pFqmuskip

\newcommand*\pFq[7][8]{
  \begingroup 
  \pFqmuskip=#1mu\relax
  \mathchardef\normalcomma=\mathcode`,
  \mathcode`\,=\string"8000
  \begingroup\lccode`\~=`\,
  \lowercase{\endgroup\let~}\pFqcomma
  {}_{#2}{#3}_{#4}{\left[\left.\genfrac..{0pt}{}{#5}{#6}\right|#7\right]}
  \endgroup
}
\newcommand{\pFqcomma}{{\normalcomma}\mskip\pFqmuskip}

\newcommand\scalemath[2]{\scalebox{#1}{\mbox{\ensuremath{\displaystyle #2}}}}

\DeclareMathOperator\arctanh{arctanh}

\allowdisplaybreaks[2]
\numberwithin{equation}{section}

\begin{document}

\vspace*{-1.5cm}
\begin{flushright}
  {\small
  ZMP-HH/21-17
  }
\end{flushright}

\vspace{1.5cm}
\begin{center}
{\LARGE
Analytic Periods via Twisted Symmetric Squares
}
\vspace{0.4cm}
\end{center}

\vspace{0.35cm}
\begin{center}
 Rafael \'Alvarez-Garc\'ia${}^{1}$,
 Lorenz Schlechter${}^{2}$
\end{center}

\vspace{0.1cm}
\begin{center} 
\emph{
${}^{1}$II. Institut f\"ur Theoretische Physik, Universit\"at Hamburg,\\
Luruper Chaussee 149, 22607 Hamburg, Germany\\
\vspace{.5cm}
${}^{2}$Institute for Theoretical Physics, Utrecht University\\
Princetonplein 5, 3584 CC Utrecht, The Netherlands } 
   \\[0.1cm] 
\vspace{0.25cm} 
\vspace{0.2cm}

\vspace{0.3cm}
\end{center}

\vspace{0.5cm}


\begin{abstract}
We study the symmetric square of Picard-Fuchs operators of genus one curves and the thereby induced generalized Clausen identities. This allows the computation of analytic expressions for the periods of all one-parameter K3 manifolds in terms of elliptic integrals. The resulting expressions are globally valid throughout the moduli space and allow the explicit inversion of the mirror map and the exact computation of distances, useful for checks of the Swampland Distance Conjecture. We comment on the generalization to multi-parameter models and provide a two-parameter example.
\end{abstract}

\clearpage

\tableofcontents


\section{Introduction}

The swampland program \cite{Vafa:2005ui} aims to distinguish those effective field theories that can be UV completed to quantum gravity theories from those that cannot. So far, this program has produced many interrelated conjectures, see \cite{Brennan:2017rbf,Palti:2019pca,vanBeest:2021lhn,Grana:2021zvf} for reviews. To test said conjectures and to better understand them a good comprehension of examples is required.

The main experimental testing grounds for these conjectures are compactifications of type II string theory or F-theory on Calabi-Yau (CY) manifolds. Compactifications to four-dimensions are described in terms of effective $\mathcal{N} = 2$ supergravity theories or $\mathcal{N} = 1$ in the presence of fluxes and orientifold planes. Many of the properties of the resulting effective theories are encoded in the periods of the CY, making their computation crucial in explicit analyses. For example, an $\mathcal{N} = 1$ supergravity theory can be completely described in terms of three objects: the superpotential $W$, the K\"ahler potential $K$ and the gauge kinetic functions. All of these quantities are expressible via the periods of the CY one compactifies on.

The usual strategy when studying flux compactifications is to obtain locally valid expressions for these objects by expanding the periods of the CY around a point in moduli space and then requiring the fluxes to be chosen in a self-consistent way such that the initially assumed approximation holds. As an example, the recently proposed constructions of exponentially small superpotentials \cite{Demirtas:2019sip,Demirtas:2020ffz,Alvarez-Garcia:2020pxd,Honma:2021klo,Demirtas:2021nlu,Demirtas:2021ote,Broeckel:2021uty,Bastian:2021hpc} develop the periods around the large complex structure point (LCS) or the conifold and drop the higher order exponentially small corrections. Globally valid expressions for the periods remove this limitation and would allow easier scans over flux vacua.

Local expressions for the periods\footnote{See \cite{Erkinger:2019umg} for similar tests of the swampland distance conjecture with a GLSM approach.} were also used in \cite{Blumenhagen:2018nts,Grimm_2018,Joshi:2019nzi} to study the swampland distance conjecture \cite{Ooguri:2006in}. Recently, the structure of the periods close to boundaries was determined in \cite{bastian2021modeling}. Near the boundary the series expansion converges slow enough as to significantly modify the distances computed by integration over trajectories traversing said regions. This was quantified in \cite{Klaewer:2021vkr} where some of the distances in \cite{Blumenhagen:2018nts} were computed exactly by exploiting heterotic-type IIA duality.

A necessary step in both computations described above is the inversion of the mirror map. The mirror map is initially defined around the LCS, characterized as the point of maximal unipotent monodromy in complex structure moduli space, and then extended to other regions of moduli space through analytic continuation, which for example defines the quantum volumes on the mirror K\"ahler moduli space. The monodromy structure around the LCS determines the well-known leading behavior of the mirror map in this regime, namely
\begin{equation}
    2\pi i t_{i}= \log(z_{i}) + a_1 z + a_2 z^2 + \cdots\,,
\end{equation}
where $t_{i}$ are the K\"ahler moduli and $z_{i}$ the complex structure moduli of the mirror manifold. The infinite series is responsible for the exponentially suppressed corrections upon inversion of the mirror map. This local expansion breaks down as we move away from the LCS and the local expression of the mirror map around other loci like the conifold is known to behave differently. As we will see in examples throughout the paper, one can combine the logarithmic term with the infinite series to form ratios of elliptic integrals that are valid throughout the whole moduli space, and that upon series expansion reproduce the expected local behavior of the mirror map.

For hypersurfaces in projective spaces and complete intersection CYs the periods have been computed in \cite{Hosono:1993qy,Hosono:1994ax} in terms of hypergeometric functions and parameter derivatives thereof. The main result of this paper is that these derivatives are in the cases of genus one curves and K3 manifolds expressible purely in terms of hypergeometric $_2F_1$ functions. The origin of these expressions lies in the fact that the Picard-Fuchs operators of K3 manifolds are symmetric squares of second order differential operators \cite{Lian:1995js,Doran:1998hm}. In general, these second order operators are not nicely behaved, but the introduction of twists allows for a unified description in terms of a basis of hypergeometric $_2F_1$ functions related to a small classified set of integer sequences. Symmetric squares have already been studied to find modular expressions for periods \cite{Lerche:1999hg, Lian:1995js,Lian:1995jv} or to systematically generate CY operators\cite{almkvist2021calabiyau}. The appearance of modular expressions in periods was first noted in \cite{Candelas:1993dm}. Similar expressions can be found in \cite{Candelas:1994hw,Lian:1994zv,Klemm:1995tj,Kachru:1995wm,Kaplunovsky:1995tm,LopesCardoso:1996zu,Kawai:1996te,LopesCardoso:1996nc}.

The periods expressed purely in terms of hypergeometric $_2F_1$ functions allow a global description of the moduli space, no longer dependent on infinite series representations. This can be used to directly test the refined swampland distance conjecture in this setup. We perform this computation for two examples, the manifolds $\mathbb{P}^{4}_{1,1,2,2,2}[8]$ and  $\mathbb{P}^{4}_{1,1,2,2,6}[12]$, finding agreement with \cite{Klaewer:2021vkr}.

This paper is organized as follows. In Section \ref{sec:periods} the idea behind the computation of the periods is discussed and applied to some examples. In Section \ref{sec:beyond-hypergeometry} we generalize this method beyond the usual hypergeometric setup. Section \ref{sec:two-moduli-models} studies a two-parameter model. Finally, Section \ref{sec:outlook} summarizes our findings and gives an outlook on possible further generalizations.

\section{Computation of the periods}
\label{sec:periods}
Our goal is to compute distances along the boundary of the complex structure moduli space of CY threefolds. For this we require the metric $g_{i\overline{j}}$ on said moduli space. The metric originates from a K\"ahler potential $K$ as
\begin{equation}
    {g_{i\overline{j}}}=\partial_i\partial_{\overline{j}} K\,,
\label{eq:metric}
\end{equation}
where
\begin{equation}
    K=-\log(i \Pi^{\dagger} \cdot \Sigma \cdot \Pi)\,.
\label{eq:Kaehler-pot}
\end{equation}
Here $\Pi$ denotes the period vector defined by
\begin{equation}
    \Pi(x)^{\alpha} = \int_{\gamma^{\alpha}} \Omega(x)\,,\quad \gamma^{\alpha} \in H_{3}(X,\mathbb{Z})\,,\quad \alpha = 0, \dots, 2 h^{2,1}(X) + 1\,,
\end{equation}
where $\Omega(x)$ is the unique holomorphic $(3,0)$-form of the CY threefold $X$, the moduli space coordinates are denoted $x$ and $\{\gamma^{\alpha}\}_{\alpha = 0,\dots,2h^{2,1}(X)+1}$ is an integral symplectic basis of 3-cycles. The symplectic pairing of the periods is given by
\begin{equation}
    \Sigma = 
    \begin{pmatrix}
    0 & \mathcal{I}\\
    -\mathcal{I} & 0\\
    \end{pmatrix}\,.
\end{equation}
Thus a knowledge of the periods is crucial for the computation of the distances. We will obtain closed forms for the periods using four ingredients:
\begin{itemize}
    \item the known formulae \cite{Hosono:1993qy, Hosono:1994ax} for the periods around the LCS point, which express the periods in terms of parameter derivatives of hypergeometric functions;
    \item Clausen's identity, reducing the problem to the solutions of second order operators;
    \item a relation between the periods originating from supersymmetry \cite{Hosono:2000eb}, which allows expressing the periods in terms of the first parameter derivatives;
    \item the so-called $\epsilon$-expansion of a family of hypergeometric functions to first order\cite{2018arXiv180804837B,2018arXiv180102428N}.
\end{itemize}
When combined, these are sufficient to compute the periods of K3 manifolds given by hypersurfaces in weighted projective spaces. These appear as fibers of Calabi-Yau threefolds. Thus, the knowledge of these periods allows the analytic computation of the distances in moduli space along deformations of these fibers. In Section 3 we will generalize this construction to more complicated geometries, including, for example, complete intersections in Grassmannian ambient spaces, by identifying a basis of integral sequences which can be resummed to closed forms. In the remainder of this section we describe the computation of the periods in detail and calculate the distances in two hypergeometric examples. To denote the complete intersections in weighted projective spaces we employ the usual matrix notation giving the degrees of the defining homogeneous polynomials in each of the projective factors
\begin{equation}
\left[
\begin{array}{c|cccc}
    \mathbb{P}^{n_{1}}_{w_1^{(1)}w_2^{(1)}\dots w_{n_1+1}^{(1)}}&d_1^{(1)}&d_2^{(1)}&\cdots&d_p^{(1)}\\
    \mathbb{P}^{n_{2}}_{w_1^{(2)}w_2^{(2)}\dots w_{n_2+1}^{(2)}}&d_1^{(2)}&d_2^{(2)}&\cdots&d_p^{(2)}\\
    \vdots&\vdots&\vdots&\ddots&\vdots\\
    \mathbb{P}^{n_{k}}_{w_1^{(k)}w_2^{(k)}\dots w_{n_k+1}^{(k)}}&d_1^{(k)}&d_2^{(k)}&\cdots&d_p^{(k)}\\
\end{array}
\right]\,.
\end{equation}
For these geometries the periods around the LCS point are well known. One can write them down explicitly in terms of derivatives of a fundamental period
\begin{equation}
    \omega_0=\sum_{n=0}^\infty c_n x^{n+\rho}\;.
\label{eq:fundamental-power-series}
\end{equation}
Here and in the following we use a multi-index notation to simplify the expressions, i.e.\ $x^n=x_1^{n_1}x_2^{n_2}\cdots x_{h^{2,1}}^{n_{h^{2,1}}}$ and $\sum_{n=0}^\infty=\sum_{n_1=0}^\infty\sum_{n_2=0}^\infty\cdots\sum_{n_{h^{2,1}}=0}^\infty$. The $\rho_{i}$ are auxiliary variables that play a role in the computation of the rest of the periods and should be set to zero at the end. The expansion coefficients are given by\cite{Hosono:1994ax}
\begin{equation}
   c_n=\frac{\prod\limits_{j=1}^p\Gamma\left(1+\sum\limits_{i=1}^k (n_i+\rho_i) d^{(i)}_j\right)
}{\prod\limits_{i=1}^k \prod\limits_{j=1}^{n_i+1}\Gamma\left(1+w_j^{(i)}(n_i+\rho_i)\right)}\,.
\end{equation}
The full basis of periods is obtained by acting with the differential operators
\begin{equation}
\begin{aligned}
    D_{1,i} &= \frac{1}{2\pi i} \partial_{\rho_{i}}\,,\\
    D_{2,i} &= \frac{1}{2} \frac{K_{ijk}}{(2\pi i)^{2}} \partial_{\rho_{j}}\partial_{\rho_{k}}\,,\\
    D_{3} &= -\frac{1}{6} \frac{K_{ijk}}{(2\pi i)^{3}} \partial_{\rho_{i}}\partial_{\rho_{j}}\partial_{\rho_{k}}\,,
\end{aligned}
\label{eq:frobenius-operators}
\end{equation}
on the fundamental period, where the $K_{ijk}$ are the classical triple intersection numbers and $i=1,\ldots,h^{2,1}$. The period vector is then 
\begin{equation}
\label{hyperbasis}
    \omega_\mathrm{LCS} =
    \left.
    \begin{pmatrix}
    \omega_{0}\\
    D_{1,i}\, \omega_{0}\\
    D_{2,i}\, \omega_{0}\\
    D_{3}\, \omega_{0}
    \end{pmatrix}
    \right\rvert_{\rho_{i} = 0}\,.
\end{equation}
This period vector represents a basis of solutions to the Picard-Fuchs system, but is not yet in an integer symplectic basis. The periods expressed in said basis are obtained by acting with a transition matrix $m$
\begin{equation}
    \Pi=m \cdot \omega_\mathrm{LCS}\,.
\end{equation}
The matrix $m$ can be fixed up to $\textrm{Sp}(2h^{2,1}+2,\mathbb{Z})$ transformations by known methods at the LCS \cite{Candelas:1993dm}.

By performing a resummation of the fundamental period we can express it as a sum of hypergeometric functions in which all of the dependence in one of the moduli space coordinates is encapsulated in the argument of the hypergeometric functions
\begin{equation}
\label{fundamental}
\omega_0=\sum_{\substack{n_i=0\\ i\neq 1}}^\infty \prod_{\substack{i=2 }}^{h^{2,1}}x_i^{n_i+\rho_i}x_1^{\rho_1}f(n_i,\rho_i)\;{}_pF_q({\vec{a}};{\vec{b}};x_1)\,.
\end{equation}
Here $f(n_i,\rho_i)$ denotes a combination of $\Gamma$ functions which does not depend on the moduli. The parameters $\vec{a}$ and $\vec{b}$ of the hypergeometric function depend on the $n_{i}$ and the $\rho_{i}$. We will be interested in one-dimensional subspaces of the moduli space where all but one modulus are fixed at the LCS point, i.e.\ $x_i=0$ for $i>1$. This immediately implies that terms in \eqref{fundamental} with $n_i>0$ will not contribute. Thus, \eqref{fundamental} simplifies to
\begin{equation}
\omega_0= \prod_{\substack{i=2}}^{h^{2,1}}x_i^{\rho_i}x_1^{\rho_1}f(\rho_i)\;{}_pF_q({\vec{a}};{\vec{b}};x_1)\;.
\end{equation}
This simplification holds as well for the $\partial_{\rho_{i}}$ derivatives since they will yield terms proportional to
\begin{equation}
    x_{i}^{n_{i}+\rho_{i}} \log(x_{i})\, f(\rho_{i}) {}_pF_q({\vec{a}};{\vec{b}};x_1) + x_{i}^{n_{i}+\rho_{i}}\, \partial_{\rho_{i}} \left(f(\rho_{i}) {}_pF_q({\vec{a}};{\vec{b}};x_1)\right)\,.
\end{equation}
We keep the $x_{i}$ strictly at $x_{i} = 0$, while we allow $x_{1}$ to take finite values along which we will integrate later on to compute distances along the boundary of the moduli space. The $\partial_{\rho_{i}} f(\rho_{i})$ give combinations of $\Gamma$ and polygamma functions that are  finite. Therefore, even if at certain $x_{1}$ we could have a competing divergence coming from the hypergeometric function or its parameter derivatives, this will happen at a zero-measure set that will not contribute to our discussion.\footnote{In general the points which have to be excluded are exactly the singular points of the manifold. For most applications one is only interested in stabilizing the moduli close to the singularities, not exactly at the singularities.} Hence, for our purposes we can say that for the $n_{i} = 0$ case the leading terms in the $x_{i}$ coordinates come from the $\log (x_{i})$ pieces without further discussion. Thus, in the regime of interest, to obtain the relevant terms we only need to determine ${}_pF_q({\vec{a}};{\vec{b}};x_1)$ and its $\rho_{1}$ derivatives. For this, the hypergeometric functions is expanded into a power series
\begin{equation}
    {}_pF_q\left({\vec{a}};{\vec{b}};x_1\right)=F_0 + \rho_1 F_1 + \rho_1^2F_2 + \cdots\,.
\end{equation}
This power series is called the $\epsilon$-expansion as the same expansions appear in the computation of scattering amplitudes in dimensional regularization \cite{Kalmykov:2008gq}. The strategy to compute the derivatives is the following. First we compute the $\epsilon$-expansion of a certain family of $_2F_1$-functions corresponding to elliptic and genus one curves. Then, using Clausen's identity and generalizations thereof, we compute the $\epsilon$-expansion of a class of $_3F_2$-functions, which will turn out to describe the geometry of K3 manifolds.

The relevant family of $_2F_1$-functions is the Legendre family of hypergometrics
\begin{equation}
\label{hypergeometric1}
    f_a(x):=\pFq{2}{F}{1}{a,1-a}{1}{x}\,.
\end{equation}
Here $0<a<1$ is a real parameter. This type of hypergeometric functions are closely related to the Legendre functions and can be expressed using them. Moreover, for $a=1/2$ the function is the complete elliptic integral $K(m)$:
\begin{equation}
    f_{1/2}=\pFq{2}{F}{1}{\frac{1}{2},\frac{1}{2}}{1}{m}=\frac{\pi}{2}K(m)\,.
\end{equation}
We will give all results of this paper in terms of either the elliptic integral or the hypergeometric $_2F_1$ functions. One could instead choose to describe them using the Legendre functions, but the hypergeometric version will allow for a unified treatment.

The hypergeometric function \eqref{hypergeometric1} describes the fundamental period $\omega_0$ of the four complete intersection elliptic curves
\begin{equation}
    \mathbb{P}^{3}_{1,1,1,1}[2,2]\,,\quad \mathbb{P}^{2}_{1,1,1}[3]\,,\quad \mathbb{P}^{2}_{1,1,2}[4]\,,\quad \mathbb{P}^{2}_{1,2,3}[6]
\end{equation}
for $a=\left\{\frac{1}{2},\frac{1}{3},\frac{1}{4},\frac{1}{6}\right\}$ respectively. Note that these are exactly the four elliptic functions appearing in Ramanujan's theory of elliptic functions to alternative bases \cite{ellipticBases}. The Picard-Fuchs system of an elliptic curve has two independent solutions. The second solution is obtained from the fundamental period by computing
\begin{equation}
\label{hypergeometric2}
    \omega_1=\frac{1}{2\pi i}\partial_{\epsilon}\; \left.\pFq{2}{F}{1}{a+\epsilon,1-a+\epsilon}{1+2\epsilon}{x}\right|_{\epsilon=0}+\log(x)f_a(x)\,.
\end{equation}
Thus, we need the $\epsilon$-expansion of a hypergeometric $_2F_1$ function to leading order in $\epsilon$. Similar problems have appeared quite often in the literature, see for example \cite{WEINZIERL2002357,Kalmykov:2006pu,Kalmykov:2008gq,Greynat:2013hox,Greynat:2014jsa,Kalmykov:2020cqz,Bytev:2020zhg}. For integer and half-integer parameters the computation is even automatized and a Mathematica package is available \cite{Huber:2007dx}. In the general case computing such expansions is a difficult task. However, for the very special form appearing in \eqref{hypergeometric2} we are in luck: a closed form can be found in terms of a harmonic series. The derivatives only with respect to the upper parameters can be found in \cite{2018arXiv180804837B} and the one with respect to the lower one in \cite{2018arXiv180102428N}. The details of the computation can be found in Appendix \ref{sec:hypergeometric-toolkit}. The result is
\begin{equation}
    \omega_1=\frac{\pi}{\sin (\pi a)} \left( f_a(x)- f_a(1-x)\right)-\left(H_{\frac{a-1}{2}}+H_{-\frac{a}{2}}+2\log (2)\right) f_a(x)\,,
\end{equation}
where $H_n$ denotes the harmonic numbers\footnote{For rational $n$ the harmonic numbers can be understood via their relation to the polygamma function, $H_n=\psi(n+1)+\gamma$\:.}. This expression simplifies drastically when one of the four explicit values of $a$ is inserted. For example, for $a=1/2$
\begin{equation}
    \omega_1=\frac{8 \log (2) K(x)}{\pi}-2 K(1-x)\,.
\end{equation}
While computing $\epsilon$-expansions directly is difficult, a surprisingly large group of periods can be computed using only this single result. For example, the four functions $f_a$ fulfill a very interesting identity due to Clausen:
\begin{equation}
    \tilde{f}_a(x):=\pFq{3}{F}{2}{a,1-a,\frac{1}{2}}{1,1}{4x(1-x)}=f_a(x)^2\;.
\end{equation}
The $_3F_2$ functions appearing in this identity are an example of the fundamental periods of one-parameter K3 surfaces. This allows us to compute the A-cycle periods of these K3s in a closed form, for details we refer again to Appendix \ref{sec:hypergeometric-toolkit}. For the B-cycle periods one also needs the second term in the $\epsilon$-expansion. But here supersymmetry comes to aid: type II string theory compactified on a K3 surface results in an $\mathcal{N}=4$ supersymmetric theory. Therefore, there are no instanton corrections to the prepotential, which results in a constraint relating the periods among themselves. This can be used to express the periods purely in terms of the first term of the $\epsilon$-expansion. For one-parameter models, such a relation can be derived directly from the properties of the hypergeometric system, with the result
\begin{equation}
\label{periodrelation2}
    \tilde{\omega}_2(x)=\frac{\tilde{\omega}_1(x)^2}{\tilde{\omega}_0(x)}\,.
\end{equation}
This is the kind of relation needed to express everything in terms of first derivatives, as only $\tilde{\omega}_2(x)$ depends on second derivatives. 
A derivation of this identity using the bilateral hypergeometric function is given in Appendix \ref{sec:hypergeometric-toolkit}. The relation also follows from the fact that the Picard-Fuchs operator for $\tilde{f}_a(x)$ is the symmetric square of the operator of $f_a(x)$. We can see this at work for the K3 periods $\tilde{\omega}$, which can be written as
\begin{align}
    \tilde{\omega}_0(x)&=\omega_0 \left( \frac{1}{2} \left(1-\sqrt{1-x}\right) \right)^2\,,\\
    \tilde{\omega}_1(x)&=\omega_0 \left( \frac{1}{2} \left(1-\sqrt{1-x}\right) \right) \omega_1 \left( \frac{1}{2} \left(1-\sqrt{1-x}\right) \right)\,,\\
    \tilde{\omega}_2(x)&=\omega_1 \left( \frac{1}{2} \left(1-\sqrt{1-x}\right) \right)^2\;.
\end{align}
In the first equation we see how Clausen's identity relates the K3 periods to those of elliptic surfaces.
Moreover, the mirror map takes the especially simple form
\begin{equation}
    t=\frac{ \tilde{\omega}_1(x)}{ \tilde{\omega}_0(x)}=\frac{\omega_1 \left( \frac{1}{2} \left(1-\sqrt{1-x}\right) \right)}{\omega_0 \left( \frac{1}{2} \left(1-\sqrt{1-x}\right) \right)}\,.
\end{equation}
Comparing this to the mirror map of the elliptic curve
\begin{equation}
    t=\frac{\omega_1(x)}{\omega_0(x)}\,,
\end{equation}
one can see that the only added complexity lies in the rational function appearing in the argument.

\subsection*{Examples}
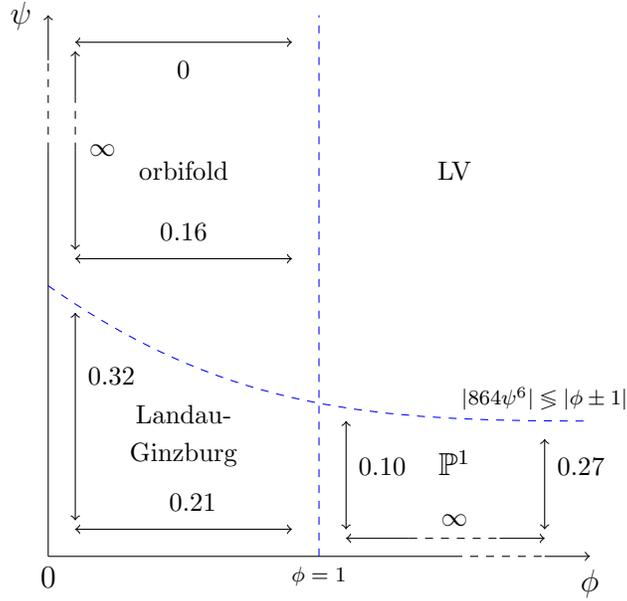
\begin{figure}[ht!]
\centering
\begin{tikzpicture}[xscale=1.2,yscale=1.2]
\node at (3,-0.3) {$\phi$};
\node at (-3.3,6) {$\psi$};
\node[align=left,below] at (-3,0) {$0$};
\node[align=center,below] at (0,0) {\scriptsize $\phi=1$};
\node[align=center,above] at (2.5,1.5) {\scriptsize $|864 \psi^6| \lessgtr |\phi \pm 1|$};

\draw[-] (-3,0) -> (-3,4.5);
\draw[dashed] (-3,4.5) -> (-3,5.5);
\draw[->] (-3,5.5) -> (-3,6);

\draw (-3,0) -> (1.5,0);
\draw[dashed] (1.5,0) -> (2.5,0);
\draw[->] (2.5,0) -> (3,0);

\draw[dashed,blue] (0,0) -> (0,6);
\coordinate (A) at (-3,3);
\coordinate (B) at (3,1.5);
\draw[dashed, blue]    (A) to[out=-35,in=180] (B);

\draw[<->] (-2.7,0.3) to (-0.3,0.3);
\draw[<->] (-2.7,0.4) to (-2.7,2.7);
\node at (-1.4,0.6) {\footnotesize 0.21};
\node at (-2.3,2) {\footnotesize 0.32};

\draw[<->] (0.3,0.3) to (0.3,1.5);
\draw[<->] (2.5,0.3) to (2.5,1.3);
\draw[<-] (0.3,0.2) to (1,0.2);
\draw[dashed] (1,0.2) to (2,0.2);
\draw[->] (2,0.2) to (2.5,0.2);
\node at (0.7,1) {\footnotesize 0.10};
\node at (2.9,1) {\footnotesize 0.27};
\node at (1.5,0.4) {\footnotesize $\infty$};

\draw[<->] (-2.7,3.3) to (-0.3,3.3);
\draw[<->] (-2.7,5.7) to (-0.3,5.7);
\draw[<-] (-2.7,3.4) to (-2.7,4.5);
\draw[dashed] (-2.7,4.5) to (-2.7,5.1);
\draw[->] (-2.7,5.1) to (-2.7,5.6);
\node at (-1.5,3.6) {\footnotesize 0.16};
\node at (-1.5,5.4) {\footnotesize $0$};
\node at (-2.4,4.5) {\footnotesize $\infty$};

\node[align=center,below] at (1.5,4.5) {\footnotesize LV};
\node[align=center,below] at (1.5,1.3) {$\mathbb P^1$};
\node[align=center,below] at (-1.5,1.8) {\footnotesize Landau- \\ \footnotesize{Ginzburg}};
\node[align=center,below] at (-1.5,4.5) {\footnotesize orbifold};
\end{tikzpicture}
\caption{\small Schematic plot of the (mirror) moduli space of $\mathbb P^{4}_{1,1,2,2,6}[12]$ taken from \cite{Blumenhagen:2018nts}. The shown distances are the ones computed in said paper. The moduli space of $\mathbb P^{4}_{1,1,2,2,2}[8]$ has an analogous structure.}
\label{fig_11226}
\end{figure}

In the rest of this section we will apply the developed methods to K3 fibered CY threefolds obtained by resolving singularities of degree eight and twelve hypersurfaces on $\mathbb{P}^{4}_{1,1,2,2,2}$ and $\mathbb{P}^{4}_{1,1,2,2,6}$ respectively and use the result to compute distances in the moduli space. Their quantum geometry has been analyzed in great detail using mirror symmetry in \cite{Candelas:1993dm,Hosono:1993qy,Hosono:1994ax}. The Fermat hypersurfaces among them are defined by the polynomials
\begin{equation}
    z_{1}^{8} + z_{2}^{8} + z_{3}^{4} + z_{4}^{4} + z_{5}^{4} = 0
\end{equation}
and
\begin{equation}
    z_{1}^{12} + z_{2}^{12} + z_{3}^{6} + z_{4}^{6} + z_{5}^{2} = 0\,.
\end{equation}
At $z_{1} = z_{2} = 0$ these geometries present a curve $C$ of $A_{1}$ singularities that when blown up lead to an exceptional divisor $E$ having the structure of a $\mathbb{P}^{1}$ fibration over $C$. The degree one polynomials generate a linear system $|L|$ which is a $\mathbb{P}^{1}$ fibration with fiber $L = \textrm{K3}$. The degree two polynomials generate a second linear system $|H| = |2L + E|$. The complexified K\"ahler cone can then be parametrized by
\begin{equation}
    K = t^{1}J_{1} + t^{2}J_{2}\,,\quad J_{1} = 2L + E\,,\quad J_{2} = L\,.
\end{equation}
Their structure as K3 fibrations can be made obvious by equivalently expressing these geometries as complete intersection Calabi-Yau manifolds (CICY)
\begin{equation}
    \mathbb{P}^{4}_{1,1,2,2,2}[8] \cong
    \left[
    \begin{array}{c|cccc}
        \mathbb{P}^{4} & 4 & 1\\
        \mathbb{P}^{1} & 0 & 2
    \end{array}
    \right]\,\quad
    \mathbb{P}^{4}_{1,1,2,2,6}[12] \cong
    \left[
    \begin{array}{c|cccc}
        \mathbb{P}^{4}_{1,1,1,1,3} & 6 & 1\\
        \mathbb{P}^{1} & 0 & 2
    \end{array}
    \right]\,,
\end{equation}
where the first factor corresponds to the K3 fiber and the second one to the $\mathbb{P}^{1}$ base. With the parametrization of the K\"ahler cone chosen for the hypersurface the $J_{i}$ coincide with the induced K\"ahler forms from the $i$-th projective factor in the CICY representation \cite{Hosono:1994ax}, and therefore we will loosely refer to $J_{1}$ and $J_{2}$ as the K\"ahler form of the fiber and the base respectively. The intersection ring was computed in \cite{Candelas:1993dm}. Since we are dealing with K3 fibrations we have $L^{2} = 0$. The non-vanishing triple intersections are given by
\begin{equation}
\begin{aligned}
    \mathbb{P}^{4}_{1,1,2,2,2}[8]\,:\quad &H^{3} = 8\,,\quad H^{2} \cdot L = 4\,,\\
    \mathbb{P}^{4}_{1,1,2,2,6}[12]\,:\quad &H^{3} = 4\,,\quad H^{2} \cdot L = 2\,.
\end{aligned}
\end{equation}
The mirror duals of these hypersurfaces can be constructed as the families of hypersurfaces defined by
\begin{equation}
    x_{1}^{8} + x_{2}^{8} + x_{3}^{4} + x_{4}^{4} + x_{5}^{4} - 8 \psi x_{1}x_{2}x_{3}x_{4}x_{5} - 2\phi x_{1}^{4}x_{2}^{4} = 0\quad \textrm{in}\, \mathbb{P}^{4}_{1,1,2,2,2}[8]
\end{equation}
and
\begin{equation}
    x_{1}^{12} + x_{2}^{12} + x_{3}^{6} + x_{4}^{6} + x_{5}^{2} - 12 \psi x_{1}x_{2}x_{3}x_{4}x_{5} - 2\phi x_{1}^{6}x_{2}^{6} = 0\quad \textrm{in}\, \mathbb{P}^{4}_{1,1,2,2,6}[12]\,,
\end{equation}
quotiented by appropriate discrete groups. The phase structure of the (mirror) moduli space is described in \cite{Aspinwall:1994ay} and depicted in terms of the $(\psi,\phi)$ variables in Figure \ref{fig_11226}.
We will be interested in computing the distance along the boundaries of moduli space. Using the language of the A-side variables with the splitting $t^{i} = \xi^{i} + i \tau^{i}$, where $\xi^{i}$ is the axion and $\tau^{i}$ its saxionic partner, this translates to keeping one of the $\tau^{i}$ strictly infinite. For example, the distance along the boundary of the $\mathbb{P}^{1}$ phase between the small $\psi$ and large $\phi$ point and the conifold point corresponds to keeping $\tau^{2} \rightarrow \infty$ while varying $t^{1}$, which greatly simplifies all the expressions involved.

In the large volume regime the $\mathcal{N} = 2$ special geometry data is encapsulated in the prepotential
\begin{equation}
    \mathcal{F} = \frac{1}{3!} K_{ijk} t^{i}t^{j}t^{k} + \frac{1}{2} a_{ij} t^{i}t^{j} + b_{i}t^{i} -\frac{1}{2}\frac{\zeta(3)\chi}{(2\pi i)^3} + \mathcal{F}_{\textrm{inst}}\,.
\label{eq:prepotential}
\end{equation}
Above
\begin{equation}
    K_{ijk} = \int_{X} J_{i} \wedge J_{j} \wedge J_{k}\,,\quad b_{i} = -\frac{1}{24} \int_{X} c_{2}(X) \wedge J_{i}
\end{equation}
and the $a_{ij}$ can be fixed modulo an irrelevant integer part by demanding that the prepotential gives periods with integer monodromies. The Euler characteristic of the manifold is denoted by $\chi(X)$ and the instanton corrections are given by $\mathcal{F}_{\textrm{inst}}$ and exponentially suppressed at large volume. From this prepotential one obtains an integral symplectic basis of periods
\begin{equation}
    \Pi = (2\mathcal{F} - t^{i}\partial_{i}\mathcal{F}, \partial_{i}\mathcal{F},1,t^{i})\,.
\end{equation}
The K\"ahler potential from which the moduli space metric stems can then be reconstructed from the periods as explained in Section \ref{sec:periods}. Taking the leading terms in $t^{1}$ one obtains
\begin{equation}
    K(\tau^{1} \rightarrow \infty) = -\log\left[ (t^{1}- \overline{t}_{1})^{3} \right]\,,
\end{equation}
while the leading terms in $t^{2}$ are
\begin{equation}
    K(\tau^{2} \rightarrow \infty) = -\log\left[ (t^{1}-\overline{t}_{1})^{2}(t^{2}-\overline{t}_{2}) \right]\,.
\end{equation}
Note that the instanton expansion breaks down in the non-geometrical phases of the moduli space and therefore demands an analytical continuation to these regions. These issues are best treated using the hypergeometric structure of the periods analyzed in Section \ref{sec:periods}. Using the procedure described there to obtain the leading terms on the mirror side, construct the integral symplectic basis and apply the mirror map will yield the same result as above as long as we only keep the leading terms in $t^{1}$ or $t^{2}$.

When keeping $\tau^{1} \rightarrow \infty$ the relevant component of the metric to compute the trajectory along the boundary of moduli space is
\begin{equation}
    g_{t^{2}\overline{t^{2}}}(\tau_{1} \rightarrow \infty) = \partial_{t^{2}}\partial_{\overline{t}^{2}} K = 0\,.
\end{equation}
This reproduces the distance computed in \cite{Blumenhagen:2018nts} along the boundary of the orbifold phase in moduli space. For the distance along the boundary of the $\mathbb{P}^{1}$ phase we keep $\tau^{2} \rightarrow \infty$ and compute the metric component
\begin{equation}
    g_{t^{1}\overline{t^{1}}}(\tau_{2} \rightarrow \infty) = \partial_{t^{1}}\partial_{\overline{t}^{1}} K = \frac{1}{2 \tau_{1}^{2}}\,.
\label{eq:metric_P1_boundary}
\end{equation}
Before we can calculate the distance along this boundary we need to compute the initial and final points of the trajectory in the $t^{1}$ variable.

\subsection{\texorpdfstring{$\mathbb{P}^4_{1,1,2,2,6}[12]$}{P4(1,1,2,2,6)[12]}}
The topological numbers corresponding to the $\mathbb{P}^4_{1,1,2,2,6}[12]$ hypersurface and appearing in its prepotential are given by
\begin{equation}
    K_{111}=4\,,\quad K_{112}=K_{121}=K_{211}=2\,,\quad b_1=\frac{13}{6}    \,,\quad b_2=1\,,\quad \chi = -252\,. 
\end{equation}
To simplify some expressions it is convenient to define a new set of coordinates in the complex structure moduli space of the mirror, related to the ones used in \eqref{eq:fundamental-power-series} and subsequent equations by
\begin{equation}
    \overline{x}_{1} := 2^{6}3^{3} x_{1} = 2^{6}3^{3} \left( -\frac{2\phi}{12^{6}\psi^{6}} \right)\,,\quad \overline{x}_{2} := 2^{2} x_{2} = 2^{2} \left( \frac{1}{4\phi^{2}} \right)\,.
\end{equation}
Above we obtained the relevant metric to compute the distance along the boundary of the $\mathbb{P}^{1}$ phase, i.e.\ for $\overline{x}_{2} = 0$. The trajectory starts on the B-side in the small $\psi$ and large $\phi$ point corresponding to $\overline{x}_{1} \rightarrow \infty$ and ends at the conifold point given by $\overline{x}_{1} = 1$, see Figure \ref{fig_11226}. This distance was numerically approximated in \cite{Blumenhagen:2018nts} to be $\Delta \overline{x}_{1} \approx 0.27$. Recently, in \cite{Klaewer:2021vkr} this value was analytically computed by exploiting heterotic-type IIA duality, obtaining the result
\begin{equation}
    \Delta \overline{x}_{1} = \frac{\log(3)}{2\sqrt{2}} \approx 0.39\,.
\end{equation}
The fundamental period of $\mathbb{P}^4_{1,1,2,2,6}[12]$ is given after resummation in the $x_{1}$ variable and keeping only the $n_{2} = 0$ terms as explained in Section \ref{sec:periods} by
\begin{equation}
    \omega_{0} = \left. x_{1}^{\rho_{1}} x_{2}^{\rho_{2}} \pFq{4}{F}{3}{1,\frac{1}{6}+\rho_{1},\frac{1}{2}+\rho_{1},\frac{5}{6}+\rho_{1}}{1+\rho_{1},1+\rho_{1},1+\rho_{1}-2\rho_{2}}{2^{6}3^{3} x_{1}}\right|_{\rho_{i}=0}\,.
\end{equation}
This expression is valid in the large complex structure regime and in other phases of the moduli space by appropriately choosing the branch of the hypergeometric function. In the large complex structure phase it can be associated to the fundamental period of the K3 fiber, which explains the appearance of a ${}_4F_3$ hypergeometric function.

The factors in the rest of the periods coming from the action of the $\partial_{\rho_{2}}$ derivative are dominated by the $\log(x_{2})$ piece, which leaves the hypergeometric function intact. Therefore, we can set $\rho_{2} = 0$ from the very start and notice that we only need to know how to compute the action of $\partial_{\rho_{1}}$ on the hypergeometric function\footnote{Here we are directly using the metric obtained above from the prepotential and therefore only need the boundary mirror map to compute the endpoints of the trajectory. Applying the procedure of Section \ref{sec:periods} to calculate the leading terms of the periods on the B-side we would also need to compute the action of $\partial^{2}_{\rho_{1}}$. A closed form for this derivative is given in Appendix \ref{sec:hypergeometric-toolkit}.} for $\rho_{1} = 0$. This corresponds to the type of hypergeometric $\epsilon$-expansion that we treat in detail in Appendix \ref{sec:hypergeometric-toolkit}, where we print an analytic closed form for it.

This allows us to obtain a closed form for the mirror map of the A-periods along the boundary. The A-periods obtained after dividing by the fundamental are, at this level of approximation,
\begin{equation}
    \begin{aligned}
    t^{1} := \frac{\omega_{1}}{\omega_{0}} &= \frac{1}{2\pi i} \left( \log(x_{1}) + \frac{\left. \frac{d}{d\rho_{1}}\,\pFq{4}{F}{3}{1,\frac{1}{6}+\rho_{1},\frac{1}{2}+\rho_{1},\frac{5}{6}+\rho_{1}}{1+\rho_{1},1+\rho_{1},1+\rho_{1}}{\overline{x}_{1}}\right|_{\rho_{1}=0}}{\pFq{4}{F}{3}{1,\frac{1}{6},\frac{1}{2},\frac{5}{6}}{1,1,1}{\overline{x}_{1}}} \right)\,,\\
    t^{2} := \frac{\omega_{2}}{\omega_{0}} &= \frac{1}{2\pi i}\log(x_{2})\,,
    \end{aligned}
\end{equation}
Exploiting the $\epsilon$-expansion of the ${}_4F_3$ hypergeometric function yields the boundary mirror map
\begin{equation}
    t^{1} = i \frac{\pFq{2}{F}{1}{\frac{1}{6},\frac{5}{6}}{1}{\frac{1}{2} \left( 1 + \sqrt{1 - \overline{x}_{1}} \right)}}{\pFq{2}{F}{1}{\frac{1}{6},\frac{5}{6}}{1}{\frac{1}{2} \left( 1 - \sqrt{1 - \overline{x}_{1}} \right)}}\,.
\label{eq:P11226_t1_mirror_map}
\end{equation}
From it we obtain the initial and final points of the trajectory, which are
\begin{equation}
    t^{1}_{i} = t^{1}(\overline{x}_{1} \rightarrow \infty) = (-1)^{2/3}\,,\quad t^{1}_{f} = t^{1}(\overline{x}_{1} = 1) = i\,.
\end{equation}
From \eqref{eq:P11226_t1_mirror_map} one can also explicitly check that all the points in the trajectory along the boundary of the $\mathbb{P}^{1}$ phase lie in the circumference of radius one in the complex plane. Parametrizing the trajectory as $t^{1}(\lambda) = (\lambda, \sqrt{1-\lambda^{2}})$ we arrive at
\begin{equation}
    \Delta \overline{x}_{1} = \int_{-\frac{1}{2}}^{0} \sqrt{\frac{d t^{1}(\lambda)}{d\lambda} \cdot \left( g_{t^{1}\overline{t^{1}}}(\tau_{2} \rightarrow \infty) \cdot \mathcal{I}_{2} \right) \cdot \frac{d t^{1}(\lambda)}{d\lambda}} = \frac{\arctanh{\left( \frac{1}{2} \right)}}{\sqrt{2}} \approx 0.39\,,
\end{equation}
which coincides with the result given in \cite{Klaewer:2021vkr}. This agreement was obvious from \eqref{eq:P11226_t1_mirror_map}, as using the fact that the inverse of the $j$-function can be written as
\begin{equation}
    j(\tau) = \frac{1728}{4\alpha (1-\alpha)} \Rightarrow \tau = i \frac{\pFq{2}{F}{1}{\frac{1}{6},\frac{5}{6}}{1}{1-\alpha}}{\pFq{2}{F}{1}{\frac{1}{6},\frac{5}{6}}{1}{\alpha}}
\end{equation}
for $\alpha = \frac{1}{2}\left( 1 - \sqrt{1 - \overline{x}_{1}} \right)$ we can rewrite the boundary mirror map as
\begin{equation}
    \overline{x}_{1} = \frac{1728}{j(t^{1})}\,,
\end{equation}
i.e.\ we have recovered the results of \cite{Candelas:1993dm,Kachru:1995fv} from the $\epsilon$-expansion of the boundary mirror map.

\subsection{\texorpdfstring{$\mathbb{P}^4_{1,1,2,2,2}[8]$}{P4(1,1,2,2,2)[8]}}

We repeat the same steps for the $\mathbb{P}^4_{1,1,2,2,2}[8]$ hypersurface. The topological numbers appearing in its prepotential are
\begin{equation}
    K_{111}=8\,,\quad K_{112}=K_{121}=K_{211}=4\,,\quad b_1=\frac{7}{3}\,,\quad b_2=1\,,\quad \chi = -168\,. 
\end{equation}
As before, we define on the mirror moduli space a new set of coordinates given by
\begin{equation}
    \overline{x}_{1} := 2^{8} x_{1} = 2^{8} \left( - \frac{2\phi}{8^{4}\psi^{4}} \right)\,,\quad \overline{x}_{2} := 2^{2} x_{2} = 2^{2} \left( \frac{1}{4 \phi^{2}} \right)\,.
\end{equation}
The fundamental period of $\mathbb{P}^4_{1,1,2,2,2}[8]$ after resummation in the $x_{1}$ variable and keeping only the $n_{2} = 0$ terms is
\begin{equation}
    \omega_{0} = \left. x_{1}^{\rho_{1}} x_{2}^{\rho_{2}} \pFq{4}{F}{3}{1,\frac{1}{4}+\rho_{1},\frac{1}{2}+\rho_{1},\frac{3}{4}+\rho_{1}}{1+\rho_{1},1+\rho_{1},1+\rho_{1}-2\rho_{2}}{2^{8} x_{1}}\right|_{\rho_{i}=0}\,.
\end{equation}
The boundary mirror map for $t^{1}$ is given by
\begin{equation}
    t^{1} := \frac{\omega_{1}}{\omega_{0}} = \frac{1}{2\pi i} \left( \log(x_{1}) + \frac{\left. \frac{d}{d\rho_{1}}\,\pFq{4}{F}{3}{1,\frac{1}{4}+\rho_{1},\frac{1}{2}+\rho_{1},\frac{3}{4}+\rho_{1}}{1+\rho_{1},1+\rho_{1},1+\rho_{1}}{\overline{x}_{1}}\right|_{\rho_{1}=0}}{\pFq{4}{F}{3}{1,\frac{1}{4},\frac{1}{2},\frac{3}{4}}{1,1,1}{\overline{x}_{1}}} \right)\,.
\end{equation}
Using the closed form for the $\epsilon$-expansion of the ${}_4F_3$ hypergeometric function given in Appendix \ref{sec:hypergeometric-toolkit} we can express the boundary mirror map as
\begin{equation}
    t^{1} = \frac{i}{2} \frac{ \sqrt{4 + 2\sqrt{2 - 2\sqrt{1-\overline{x}_{1}}}}}{\sqrt{2 + \sqrt{2}\sqrt{1 + \sqrt{1 - \overline{x}_{1}}}}} \frac{K\left( \frac{2\sqrt{2} \sqrt{1 + \sqrt{1 - \overline{x}_{1}}}}{2 + \sqrt{2} \sqrt{1 + \sqrt{1 - \overline{x}_{1}}}} \right)}{K\left( \frac{2\sqrt{2 - 2\sqrt{1 - \overline{x}_{1}}}}{2 + \sqrt{2 - 2\sqrt{1 - \overline{x}_{1}}}} \right)}\,,
\label{eq:P11222_t1_mirror_map}
\end{equation}
where $K(m) = \frac{\pi}{2} {}_2F_{1} \left( \frac{1}{2},\frac{1}{2};1;m \right)$ is the complete elliptic integral of the first kind. Note that the infinite sum combines with the logarithm to form the elliptic integrals. Of course, if the right hand side of \eqref{eq:P11222_t1_mirror_map} is expanded into a power series around $\ov{x}_1=0$ the logarithm reappears and one obtains the usual form of the mirror map.

The initial and final points of the trajectory are
\begin{equation}
    t^{1}_{i} = t^{1}(\overline{x}_{1} \rightarrow \infty) = -\frac{1}{2} + \frac{i}{2}\,,\quad t^{1}_{f} = t^{1}(\overline{x}_{1} = 1) = \frac{i}{\sqrt{2}}\,.
\end{equation}
From \eqref{eq:P11222_t1_mirror_map} one can see that all the points in the trajectory lie in the circumference of radius $\frac{1}{\sqrt{2}}$ in the complex plane. Parametrizing the trajectory as $t^{1}(\lambda) = \left(\lambda, \sqrt{\frac{1-2\lambda^{2}}{2}}\right)$ we arrive at
\begin{equation}
    \Delta \overline{x}_{1} = \int_{-\frac{1}{2}}^{0} \sqrt{\frac{d t^{1}(\lambda)}{d\lambda} \cdot \left( g_{t^{1}\overline{t^{1}}}(\tau_{2} \rightarrow \infty) \cdot \mathcal{I}_{2} \right) \cdot \frac{d t^{1}(\lambda)}{d\lambda}} = \frac{\arctanh{\left( \frac{1}{\sqrt{2}} \right)}}{\sqrt{2}} \approx 0.62\,,
\end{equation}
which coincides with the result given in \cite{Klaewer:2021vkr}. There, the boundary mirror map was given in terms of the Hauptmodul $j_{2}^{+}$ by considerations of modularity as
\begin{equation}
    \overline{x}_{1} = \frac{256}{j_{2}^{+}(t^{1})}\,.
\end{equation}
An explicit expression for $j_{2}^{+}$ is given by \cite{Harnad:1998hh}
\begin{equation}
    j_{2}^{+}(t^{1}) = 2^{4} \frac{(\theta_{3}(q_{t^{1}})^{4} + \theta_{4}(q_{t^{1}})^{4})^{4}}{\theta_{2}(q_{t^{1}})^{8} \theta_{3}(q_{t^{1}})^{4} \theta_{4}(q_{t^{1}})^{4}}\,,
\end{equation}
where $q_{t^{1}} = \exp{(i \pi t^{1})}$. Thanks to the hypergeometric treatment of the periods we have automatically obtained the exact inverse \eqref{eq:P11222_t1_mirror_map} of this relation.

\section{Beyond hypergeometry}
\label{sec:beyond-hypergeometry}

Up to now, we focused on the computation of periods expressible via hypergeometric functions. This allows for the computation of the periods for one-parameter subspaces in toric CICYs. If one leaves this setup, more complicated GKZ systems appear. In this section we will venture beyond the hypergeometric case to obtain the periods of non-toric surfaces as well. This includes all one-parameter K3 surfaces of \cite{Lian:1995js} and the fiber in all examples of \cite{Knapp:2021vkm} when taking the limit $z_2\rightarrow 0$ (see \cite{Kimura:2016crs,Kimura:2019bzv} for related constructions involving Grassmannians). Moreover, all Picard-Fuchs operators of one-parameter Fano threefolds can be computed this way. This includes, for example, sections of the Grassmannian $G(2,6)$ by a codimension 5 plane. One-parameter Fano threefolds are classified, for the complete list see \cite{golyshev2007classification}.  

With the periods at hand one could repeat the analysis of Section \ref{sec:periods} for the distances along the boundary of the moduli space for CY threefolds admitting a fibration by these surfaces. The list \cite{Lian:1995js} contains examples with increasing values of $N$ for the relevant congruence subgroup $\Gamma_{0}(N)^{+}$, going up to $N = 30$, which would allow for an analysis similar to that of \cite{Klaewer:2021vkr}.

The hypergeometric differential operators for elliptic curves are of the form
\begin{equation}
\mathcal{L}_{\textrm{ell}} = \theta^2+C x (\theta+a_1)(\theta+a_2)\,,
\end{equation}
or for the one-parameter K3 case
\begin{equation}
\mathcal{L}_{\textrm{1-K3}} = \theta^3+C x (\theta+a_1)(\theta+a_2)(\theta+a_3)\,,
\end{equation}
where we have used the logarithmic derivative $\theta = x \partial_{x}$. The general second order operator equation can be written as
\begin{equation}
\tilde{\mathcal{L}}=\theta^2+\sum\limits_{n=1}^m x^n P_n(\theta)\,,
\end{equation}
for some fixed $m$, known as the degree of the operator, and polynomials $P_n$ of  degree 2. In this paper we will only slightly venture beyond the hypergeometric case, where $m=1$, by taking $m=2$. The interesting operators of this type have been studied by \cite{2005math......7430A,Almkvist:2004kj,zagier2009,almkvist_van_straten_zudilin_2011,2013arXiv1304.5434B,2013arXiv1310.6658A,2021arXiv210211839G,2021arXiv210308651A}. In those papers the differential equation
\begin{equation}
\label{diffeqzagier}
    \left[ \theta^2-x(A\theta^2+A\theta+B)+x^2C(\theta+1)^2 \right] f(x)=0\,,
\end{equation}
or the corresponding recursion (known as an Ap\'ery-like recursion)
\begin{equation}
    n^2 u_n- (A (n-1)^2+A (n-1)+B)u_{n-1}+C(n-1)^2u_{n-2}=0\,,
\label{eq:apery-like-recursion}
\end{equation}
for the coefficients of a power series Ansatz $f(x) = \sum_{n=0}^{\infty} u_{n} x^{n}$ were considered. Here A, B and C are integer parameters. The found solutions with integer coefficients are all either polynomials, Legendre functions or hypergeometric functions. In addition to these infinite families of solutions, there are six sporadic solutions, denoted A-F by Zagier, and an additional hypergeometric solution that does not fall under the one-parameter family of hypergeometric solutions, denoted G. We list all solutions in Table \ref{solutionstable}.

\begin{table}[h!]
    \centering
    {\tabulinesep=2mm
    \begin{tabu}{|c|c|c|c|c|}
    \hline
    A & B & C & Name & Solution\\
    \hhline{=====}
    16 & 4 & 0 & & $\pFq{2}{F}{1}{\frac{1}{2},\frac{1}{2}}{1}{x}$\\
    \hline
    27 & 6 & 0 & & $\pFq{2}{F}{1}{\frac{1}{3},\frac{2}{3}}{1}{x}$\\
    \hline
    64 & 12 & 0 & & $\pFq{2}{F}{1}{\frac{1}{4},\frac{3}{4}}{1}{x}$\\
    \hline
    432 & 60 & 0 & & $\pFq{2}{F}{1}{\frac{1}{6},\frac{5}{6}}{1}{x}$\\
    \hline
    0 & 0 & -16 & G & $\frac{1}{2}\left(\pFq{2}{F}{1}{\frac{1}{2},\frac{1}{2}}{1}{16x}+\pFq{2}{F}{1}{\frac{1}{2},\frac{1}{2}}{1}{-16x}\right)$\\
    \hline
    7 & 2 & -8 & A & $\frac{1}{1-2x}\; \pFq{2}{F}{1}{\frac{1}{3},\frac{2}{3}}{1}{27x^2/(1-2x)^3}$\\
    \hline
    \multirow{2}{*}{9} & \multirow{2}{*}{3} & \multirow{2}{*}{27} & \multirow{2}{*}{B} & $\frac{1}{1-3x}\; \pFq{2}{F}{1}{\frac{1}{3},\frac{2}{3}}{1}{-27x^3/(1-3x)^3}$\\
    & & & & $=\; \pFq{2}{F}{1}{\frac{1}{3},\frac{1}{3}}{1}{27x(1-9x+27x^2)}$\\
    \hline
    \multirow{2}{*}{12} & \multirow{2}{*}{4} & \multirow{2}{*}{32} & \multirow{2}{*}{E} & $\frac{1}{1-4x}\; \pFq{2}{F}{1}{\frac{1}{2},\frac{1}{2}}{1}{16x^2/(1-4x)^2}$\\
    & & & & $=\; \pFq{2}{F}{1}{\frac{1}{2},\frac{1}{2}}{1}{16x(1-4x)}$ \\
    \hline
    17 & 6 & 72 & F & $\frac{1}{1-6x}\; \pFq{2}{F}{1}{\frac{1}{3},\frac{2}{3}}{1}{-27x^3(8x-1)/(1-6x)^3}$\\
    \hline
    10 & 3 & 9 & C & $\frac{\sqrt{2}}{p_1^{1/2}}\; \pFq{2}{F}{1}{\frac{1}{2},\frac{1}{2}}{1}{64x^3/p_1^2}$\\
    \hline
    11 & 3 & -1 & D & $\frac{1}{p_2^{1/4}}\; \pFq{2}{F}{1}{\frac{1}{12},\frac{5}{12}}{1}{1728x^5(1-11x-x^2)/(p_2)^3)}$\\
    \hline
    2 & $d^2$-$d$+1 & 1 & & $\frac{1}{(1-x)^{1-d}}\pFq{2}{F}{1}{d,d}{1}{x}$\\
    \hline
    -1 & $d^2$-$d$ & 0 & & $\pFq{2}{F}{1}{d,1-d}{1}{x}$\\
    \hline
    \end{tabu}
    }
    \caption{Solutions to Ap\'ery-like recursions \eqref{eq:apery-like-recursion}. Note that the $_2F_1\left(\frac{1}{12},\frac{5}{12};1;y\right)$ function in case D is related to $_2F_1\left(\frac{1}{6},\frac{5}{6};1;f(y)\right)$ by a rational transformation. Beukers \cite{Beukers2002} gives solutions for all of the cases A-F in terms of  $_2F_1\left(\frac{1}{12},\frac{5}{12};1;x\right)$ functions which are related to our solutions by M\"obius transformations. }
    \label{solutionstable}
\end{table}

We have defined
\begin{equation}
    \begin{aligned}
    p_1&=1-6x-3x^2+\sqrt{(1-x)^3(1-9x)}\\
    p_2&=1-12x+14x^2+12x^3+x^4
    \end{aligned}
\end{equation}
to shorten the expressions in the table. The Legendrian cases as well as the polynomial ones are special cases of hypergeometric functions.  Thus, all cases reduce to hypergeometric functions of the form $_2F_1\left(a,1-a;1;x\right)$ and rational functions. The $\epsilon$-expansion of these function is discussed in Appendix \ref{sec:hypergeometric-toolkit}. Four of the sporadic cases, A, B, E and F, seem to follow a similar structure and case G is a sum of two hypergeometric functions, while the remaining two cases are more complicated. Especially case D appears a priori complicated, but noticing that
\begin{equation}
    \pFq{2}{F}{1}{\frac{1}{12},\frac{5}{12}}{1}{\frac{1728}{j(t)}}=E_4^{1/4}(t)
\end{equation}
and that the $j$-function of the corresponding elliptic curve is given by 
\begin{equation}
j(t) = \frac{(t^4-12t^3+ 14t^2+ 12t+ 1)^3}{t^5(t^2-11t-1)}\,,
\end{equation}
explains the rather complicated structure in terms of the modularity of the underlying curve.

We will see that the first four entries are related to the fundamental periods of K3 manifolds constructed as complete intersections in toric varieties \cite{Hosono:1994ax}. Constructing the other cases is more involved, e.g.\ case D appears as a fiber in all examples of \cite{Knapp:2021vkm} when taking the limit $z_2\rightarrow 0$. Moreover, the sporadic solutions appear in the Picard-Fuchs systems of the 17 one-parameter Fano threefolds \cite{golyshev2007classification}.

All solutions $f(x)$ of \eqref{diffeqzagier} enjoy a generalization of Clausen's identity in the form of a twisted symmetric square, i.e.\ their square is related to the solutions of the differential equation
\begin{equation}
\label{diffeq}
    \left[ \theta^3-2x(2\theta+1)(A\theta^2+A\theta+B)+4C x^2(\theta+1)(2\theta+1)(2\theta+3) \right] \tilde{f}(x)=0\,,
\end{equation}
as
\begin{equation}
\label{ClausenGen}
    f(x)^2=\frac{1}{1-C x^2}\tilde{f}\left(\frac{1-Ax+C x^2}{(1-C x^2)^2}\right)\,.
\end{equation}
The geometric origin of the term twisted symmetric square is the following: The K3 manifolds can be written as twisted products of two elliptic curves $E_t\times E_{\imath(t)}$, where $\imath$ denotes an Atkin–Lehner involution. Details of this construction can be found in \cite{Peters1986,peters1989pencil,golyshev2007classification,Almkvist2011GeneralizationsOC}. For this paper the important fact is that the twisted symmetric squares have similar properties to those of the usual symmetric squares, especially the relations between the solution spaces of the operators. 
There exist several twists for each second order operator. For example, the solutions to another third order operator
\begin{equation}
\label{diffeq2}
    \left[ \theta^3-2x(2\theta+1)(\hat{A}\theta^2+\hat{A}\theta+\hat{B})+4 x^2\hat{C}(\theta+1)^3 \right] \tilde{f}(x)=0
\end{equation}
are obtained via
\begin{equation}
\label{ClausenGen2}
    f(x)^2=\frac{1}{1-Ax+Cx^2}\tilde{f}\left(\frac{-x}{1-Ax+Cx^2}\right)\,,
\end{equation}
with $A = 2\hat{A}$, $B = \hat{A} - \hat{B}$ and $C = \hat{A}^{2} - \hat{C}$.

When looking through the list of diagonal K3 Picard-Fuchs operators in \cite{Lian:1995js}, one finds that the operators in \eqref{diffeq} and \eqref{diffeq2} are not sufficient to cover all cases. Indeed, allowing for another parameter $D$ in the recursion, three more sequences were found in \cite{Cooper2012}. There the operator
\begin{equation}
\label{diffeq3}
    \theta^3-2x(2\theta+1)(A\theta^2+A\theta+B)+4 x^2(C(\theta+1)^3+D(\theta+1))
\end{equation}
was studied. The additional solutions correspond to one of the two types, either
\begin{equation}
    \theta^3-2x(2\theta+1)(A\theta^2+A\theta+B)+A x^2(\theta+1)(3\theta+2)(4\theta+3)\,,
\end{equation}
or
\begin{equation}
    \theta^3-2x(2\theta+1)(A\theta^2+A\theta+B)+A x^2(\theta+1)(4\theta+3)(4\theta+5)\,.
\end{equation}
Together with the symmetric squares of the operators in Table \ref{solutionstable} these give solutions to all operators listed in \cite{Lian:1995js}. Note that the exact parameter combination for the $\Gamma_0(2)^-$ cases does not appear. The reason for this is that in this case the parameters take values outside the parameter range considered in \cite{Cooper2012}, see \mbox{Table \ref{solutionstable2}}. While the combination gives an integer sequence like in the other cases, we could not identify a generating function for this case. However, there is also no known geometric construction associated to it, so we will simply ignore it. These represent, as in the other cases, also symmetric squares of quadratic operators.
\begin{table}[h!]
    \centering
    {\tabulinesep=2mm
    \begin{tabu}{|c|c|c|c|c|}
    \hline
    A & B & C & D & Solution\\
    \hhline{=====}
    6 & 2 & 64 & $-4$ & $\frac{\pFq{2}{F}{1}{\frac{1}{8},\frac{3}{8}}{1}{\frac{4 \left(\sqrt{4 x+1}-\sqrt{1-16 x}\right)^5 \left(\sqrt{1-16 x}+\sqrt{4 x+1}\right)}{5 \left(2 \sqrt{1-16 x}+3 \sqrt{4 x+1}\right)^4}}^2}{\frac{2}{5} \sqrt{1-16 x}+3 \sqrt{4 x+1}}$\\
    \hline
    \multirow{2}{*}{13} & \multirow{2}{*}{4} & \multirow{2}{*}{$-27$} & \multirow{2}{*}{3} & $\sqrt{\frac{\sqrt{-27 x^2-26 x+1}-13 x+1}{8 x^2+(1-8 x) \sqrt{-27 x^2-26 x+1}-21 x+1}}$\hphantom{aaaaaaaaaaaaaaaa}\\
     & & & & \hphantom{aaaaa}$\cdot\; \pFq{2}{F}{1}{\frac{1}{12},\frac{5}{12}}{1}{\frac{13824 x^7}{\left(8 x^2-21 x+(1-8 x) \sqrt{-27 x^2-26 x+1}+1\right)^3}}^2$\\
    \hline
    14 & 6 & 192 & $-12$ & $\frac{\, \pFq{2}{F}{1}{\frac{1}{8},\frac{3}{8}}{1}{\frac{256 x^3}{(1-12 x)^2}}^2}{\sqrt{1-12 x}}+1$\\
    \hline
    27 & 12 & 729 & $-81$ & ?\\
    \hline
    \end{tabu}
    }
    \caption{Solutions to Ap\'ery-like recursions \eqref{eq:apery-like-recursion} found by Cooper\cite{Cooper2012}. The appearing hypergeometric functions can be transformed into the $_2F_1(a,1-a;1;x)$ type. The last row follows from the $\Gamma_0(2)^-$ case in \cite{Lian:1995js}. This is the only case where a closed form is unknown.}
    \label{solutionstable2}
\end{table}

If we know the $\epsilon$-expansion of all $f(x)$, we are also able to compute the $\epsilon$-expansion of the $\tilde{f}(x)$. The latter fulfill the Picard-Fuchs equations of certain diagonal K3 manifolds. Thus, it is possible to use the methods of this paper to give closed forms for these manifolds.
As an example we take the sequence E. This sequence follows the recursion
\begin{equation}
    n^2 u_n- (12 (n-1)^2+12 (n-1)+4)u_{n-1}+32(n-1)^2u_{n-2}=0\,,
\end{equation}
which has the solution \cite{Gorodetsky2021NewRF}
\begin{equation}
    u_n=\sum\limits_{k=0}^n\binom{n}{k}\binom{2k}{k}\binom{2n-2k}{n-k}=\binom{2n}{n}\;\pFq{2}{F}{1}{\frac{1}{2},-n,-n}{1,\frac{1}{2}-n}{-1}\,.
\end{equation}
The generating function of this sequence, which will play the role of the fundamental period in our case, is given by
\begin{equation}
    \omega_0=\sum_{n=0}^\infty u_n x^n=\;\pFq{2}{F}{1}{\frac{1}{2},\frac{1}{2}}{1}{16x(1-4x)}\,.
\end{equation}
We only need the first term in $\epsilon$-expansion of this function, i.e.\ we need to compute
\begin{align}
    \omega_1&=\partial_\epsilon\sum_{n=0}^\infty u_{n+\epsilon} x^{n+\epsilon}\nonumber|_{\epsilon=0}\\
    &=\left.\partial_\epsilon\left((16x(1-4x))^\epsilon\frac{\Gamma\left[\frac{1}{2}+\epsilon\right]^2}{\pi \Gamma[1+\epsilon]^2}\pFq{3}{F}{2}{\frac{1}{2}+\epsilon,\frac{1}{2}+\epsilon,1}{1+\epsilon,1+\epsilon}{16x(1-4x)}\right)\right|_{\epsilon=0}\nonumber\\
    &+\log(x)\;\pFq{2}{F}{1}{\frac{1}{2},\frac{1}{2}}{1}{16x(1-4x)} \,.
\end{align}
The $\epsilon$-expansion of the $\Gamma$ functions and the rational function are easily computed, and the derivatives with respect to $\epsilon$ of the $_3F_2$ function reduce to derivatives of $_2F_1$ functions of the type discussed in Appendix \ref{sec:hypergeometric-toolkit}:
\begin{equation*}
    \partial_\epsilon\;\pFq{3}{F}{2}{\frac{1}{2}+\epsilon,\frac{1}{2}+\epsilon,1}{1+\epsilon,1+\epsilon}{16x(1-4x)}=\partial_\epsilon\; \pFq{2}{F}{1}{\frac{1}{2}+\epsilon,\frac{1}{2}+\epsilon}{1+2\epsilon}{16x(1-4x)}\,.
\end{equation*}
As the next example we take the sequence B. This has the generating function 
\begin{equation}
    \omega_0=\sum_{n=0}^\infty u_n x^n=\;\pFq{2}{F}{1}{\frac{1}{3},\frac{1}{3}}{1}{27x(1-9x+27x^2)}\,.
\end{equation}
Thus, for the second period we need to compute
\begin{align}
    \omega_1&=\partial_\epsilon\sum_{n=0}^\infty u_{n+\epsilon} x^{n+\epsilon}\nonumber\\
    &=\partial_\epsilon\left((27x(1-9x+27x^2))^\epsilon\frac{\Gamma[1/3+\epsilon]^2}{\pi \Gamma[1+\epsilon]^2}\pFq{2}{F}{1}{\frac{1}{3}+\epsilon,\frac{1}{3}+\epsilon}{1+2\epsilon}{27x(1-9x+27x^2)}\right)\nonumber\\
    &+\log(x)\;\pFq{2}{F}{1}{\frac{1}{3},\frac{1}{3}}{1}{27x(1-9x+27x^2)}\,.
\end{align}
Now we encounter a new type of $\epsilon$-expansion, that of hypergeometric functions of the type ${}_2F_1(a,a;1;x)$. But these are related to the Legendrian type via a rational transformation:
\begin{equation}
    \pFq{2}{F}{1}{a,a}{c}{x}=\frac{1}{(1-x)^a}\pFq{2}{F}{1}{a,c-a}{c}{\frac{x}{x-1}}\,.
\end{equation}
Thus, for the $\epsilon$-expansion we are interested in we have the relation
\begin{equation}
    \pFq{2}{F}{1}{a+\epsilon,a+\epsilon}{1+2\epsilon}{x}=\frac{1}{(1-x)^{(a+2\epsilon)}}\pFq{2}{F}{1}{a+2\epsilon,1-a+2\epsilon}{1+2\epsilon}{\frac{x}{x-1}}\,.
\end{equation}
This reduces the computation again to the known expansions. 

\section{Two-moduli models}
\label{sec:two-moduli-models}

So far we have been concerned with one-parameter models, but most of the machinery carries over to multi-parameter models. In this section we will discuss the generalization to multi-parameter models and work out explicitly the example of the two-parameter K3 fiber of the three-parameter CY $\mathbb{P}^4_{1,1,2,8,12}[24]$.

The two crucial ingredients necessary for the construction to work are the existence of a relation between the periods which allows the elimination of the second parameter derivative and the existence of a Clausen-like identity which expresses the fundamental period as products of $_2F_1$ functions.

The required relation between the periods was observed in \cite{Hosono:2000eb} to be
\begin{equation}
\label{periodrelation}
    D_2\, \omega_0 = \frac{1}{2}\frac{\sum_{a,b}K_{a,b}\;D_{1,a}\,\omega_0\;D_{1,b}\,\omega_0}{\omega_0} - 1\,,
\end{equation}
where $K_{a,b}$ are the classical intersection numbers $K_{a,b} = \int_{X} J_{a} \wedge J_{b}$ and the differential operators are defined as in \eqref{eq:frobenius-operators} but substituting $K_{ijk}$ for $K_{ab}$. 

This equation is the multi-moduli generalization of equation \eqref{periodrelation2}. The relation allows, independent of the number of moduli, the elimination of the $\epsilon^2$ terms in the $\epsilon$-expansion in favor of the first $\epsilon$-derivatives. The origin of this relation is the absence of instanton corrections in an $\mathcal{N}=4$ supersymmetric theory. The relation \eqref{periodrelation} is exactly of the form to ensure the absence of such corrections.

The second relation required is a Clausen-like formula to express the fundamental period as products of $_2F_1$ functions. A two-parameter example of such a formula is given by the Picard-Fuchs system \cite{Lian:1995jv}
\begin{align}
\label{PFsystem}
    \mathcal{L}_1& = \theta_x\left(\theta_x-2\theta_z\right)-x\left(\theta_x+\frac{1}{6}\right)\left(\theta_x+\frac{5}{6}\right)\,,\\
    \mathcal{L}_2 & =\theta_z^2-z\left(2\theta_z-\theta_x+1\right)\left(2\theta_z-\theta_x\right)\,.
\end{align}
This is the Picard-Fuchs system of $\mathbb{P}^4_{1,1,2,8,12}[24]$ in the limit $y\rightarrow 0$, where it reduces to the Picard-Fuchs system of the K3 surface $\mathbb{P}^{3}_{1,1,4,6}[12]$. The solutions of this system can be expressed purely in terms of those of a further reduction along $z \rightarrow 0$, where we obtain the Picard-Fuchs system
\begin{equation}
    \mathcal{L} = \lim_{z\rightarrow 0} \mathcal{L}_1=\theta_x^2-x\left(\theta_x+\frac{1}{6}\right)\left(\theta_x+\frac{5}{6}\right)
\end{equation}
for the family of elliptic curves $\mathbb{P}^{2}_{1,2,3}[6]$. Said solutions are explicitly given by
\begin{align}
    \omega_0&=\pFq{2}{F}{1}{\frac{1}{6},\frac{5}{6}}{1}{x}\,,\\
    \omega_1&=-2\pi\;\pFq{2}{F}{1}{\frac{1}{6},\frac{5}{6}}{1}{1-x}+\log(432)\;\pFq{2}{F}{1}{\frac{1}{6},\frac{5}{6}}{1}{x}\,.
\end{align}
The combinations $\omega_0(S)\,\omega_0(R)$, $\omega_1(S)\,\omega_0(R)$ and $\omega_1(S)\,\omega_1(R)$ are all annihilated by $\mathcal{L}_1$ and $\mathcal{L}_2$, where $S$ and $R$ are algebraic functions given by the solutions to the system
\begin{align}
   & R+S-864RS-x=0\,,\\
   & RS-(1-432R)(1-432S)-x^2z=0\,.
\end{align}
These equations can be solved algebraically using only radicals, but the expressions are too convoluted to represent here. Thus, the periods of this two-parameter K3 are completely expressible using only hypergeometric functions. Note that the analytic continuation of $_2F_1$ functions is well understood, so the resulting expressions are valid globally in moduli space. 

The system \eqref{PFsystem} is of course not the only system of this sort. It is part of a family: for each of the four elliptic hypergeometric systems there exists a two-parameter K3 related as above. Similar constructions can be carried out for the sporadic solutions, albeit the algebraic functions appearing in the transformation become rather cumbersome. The proofs of these statements can be found in \cite{Lian:1995jv} and are based on a twisted version of the symmetric square.

\section{Summary and outlook}
\label{sec:outlook}

The periods of K3 manifolds and Fano threefolds are obtained as solutions of the appropriate Picard-Fuchs operators. Obtaining a complete basis of periods involves taking the $\epsilon$-expansion of the fundamental period up to second order. By noting that these differential operators are the (possibly twisted) symmetric squares of lower order Picard-Fuchs operators one obtains generalized Clausen identities that allow us to express the periods of the K3 and Fano surfaces in terms of the periods of the elliptic surfaces. The complete intersection elliptic surfaces correspond to the solutions of the hypergeometric Apéry-like sequences, while the curves with four singularities correspond to the sporadic solutions. In Appendix \ref{sec:hypergeometric-toolkit} we review how the first $\epsilon$-derivative of these solutions to the Apéry-like recursions is computed. Therefore, the generalized Clausen identities provide us with most of the periods of the surfaces. By taking into account that the $\mathcal{N} = 4$ supersymmetry in the resulting compactifications demands the absence of instanton corrections an extra constraint on the periods emerges that allows for the computation of the remaining second $\epsilon$-derivative. This was checked explicitly for the hypergeometric case in Appendix \ref{sec:hypergeometric-toolkit}, while the more general relation valid also for multi-parameter models is discussed in Section \ref{sec:two-moduli-models}.

The expressions for the periods resulting from the previously described analysis are globally valid over the moduli space, allowing, for example, for computations of distances that do not suffer from the convergence problems that local expressions can present. As an application, we computed exact analytical distances along the boundary of the moduli space for some K3 fibrations, obtaining agreement with the results of \cite{Klaewer:2021vkr}. Since our computation yields directly the periods on the K\"ahler side, we obtained as a byproduct the exact inverse of the mirror map used in \cite{Klaewer:2021vkr} by considerations of modularity.

A few questions remain open for further investigation. First, it is interesting to note that Beauville's classification of elliptic curves with four singularities into six families \cite{beauville1982} is not exactly in one-to-one correspondence with the six sporadic solutions of Zagier. Rather, the map is as given in Table \ref{tab:BeauvilleClass} \cite{zagier2009}.
\begin{table}[h!]
    \centering
    {\tabulinesep=2mm
    \begin{tabu}{|c|c|}
    \hline
    Beauville type & Sporadic sequence\\
    \hline
    I, VI & B\\
    \hline
    II, V & E, G\\
    \hline
    III & D\\
    \hline
    IV & A, C, F\\
    \hline
    \end{tabu}
    }
    \caption{Correspondence between Beauville's classification of elliptic curves with four singularities and the sporadic solutions of Zagier.}
    \label{tab:BeauvilleClass}
\end{table}
All known cases of diagonal K3 surfaces include all different curves of Beauville type IV, i.e.\ they are related to the sporadic sequences A, C, and F. The Pfaffian GLSMs contain fibers of type III that are in turn related to the sporadic sequence D.

The minimal number of singular fibers in a stable family of elliptic surfaces over $\mathbb{P}^{1}$ is four, with the classification just discussed above. It would be interesting to contemplate cases presenting a higher number of singularities and perform a similar analysis.

A natural next step beyond the cases of elliptic curves and K3 manifolds would be that of Calabi-Yau threefolds. In the same way that we took symmetric squares of hypergeometric $_2F_1$ functions in order to move to the case of surfaces, we could take symmetric cubes to move to the case of threefolds. This is unfruitful as no CY threefold corresponding to the resulting operators is known in the literature. However, thus far only non-twisted symmetric cubes have been studied. It could therefore be that beyond these simpler cases an operator associated to a CY threefold can be obtained. In the case of elliptic curves and of the K3 surfaces the underlying modularity properties, although not exploited directly in this work, could very well be responsible for the simple expressions obtained after the computations. Since no analogue for the modularity of the elliptic curves is known in the CY threefold case this difference could constitute an obstruction to the application of the same techniques for this case in order to obtain global closed expressions for the periods. Also, no analogue of the period relations are known as there is less supersymmetry available. However, there still exist differential relations between the periods originating from the boundary Hodge structures \cite{bastian2021modeling} which could be used instead.

Finally, in the present paper we have only focused our attention on families of second order differential operators of degree one, the hypergeometric case, and of degree two, the beyond hypergeometric case. These are well studied and could be considered classified. The second order operators of degree three and above are less explored and it would be interesting to carry out a similar analysis for them. We plan on studying these cases and their K3 analogues in a future paper.


\vspace{0.5cm}

\noindent
\subsubsection*{Acknowledgments}
The authors would like to thank the Max-Planck-Institut f\"ur Physik for hospitality during part of this work. They would also like to thank Thomas Grimm, Damian van de Heisteeg, Daniel Kl\"awer, Timo Weigand and Sam Woldringh for comments on the draft. The work of RAG is supported in part by the Deutsche Forschungsgemeinschaft under Germany's Excellence Strategy EXC 2121 Quantum Universe 390833306.

\clearpage
\appendix

\section{Hypergeometric toolkit}
\label{sec:hypergeometric-toolkit}

In this section we will set up some mathematical identities for hypergeometric functions and their parameter derivatives needed for the computation of the periods as described in Section \ref{sec:periods}.

The generalized hypergeometric function is defined as
\begin{equation}
    \pFq{p}{F}{q}{a_{1},\dots,a_{p}}{b_{1},\dots,b_{q}}{x} := \frac{\prod\limits_{i=1}^q\Gamma[b_i]}{\prod\limits_{i=1}^p\Gamma[a_i]}\sum\limits_{n=0}^\infty \frac{\prod\limits_{i=1}^p\Gamma[a_i+n]}{\prod\limits_{i=1}^q\Gamma[b_i+n]}\frac{x^n}{\Gamma[n+1]}\;.
\end{equation}
These functions are defined by two parameter vectors $\vec{a},\,\vec{b}\in \mathbb{R}^n$ of respective lengths $p$ and $q$. For the case we are interested in $p=q+1$, i.e.\ there are as many $\Gamma$ functions in the numerator as in the denominator. Moreover, $p$ will be related to the dimension of the space we are interested in, i.e.\ a ${}_3F_2$ will describe the fundamental period of a complex one-dimensional space, a ${}_4F_3$ corresponds to a two-dimensional space and finally a ${}_5F_4$ will describe a CY threefold.\footnote{Here we refer to the $_pF_{q}$ whose parameter derivatives we need to compute in order to obtain the periods beyond the fundamental one, c.f.\ \eqref{hyperbasis}. After setting $\rho_{i} = 0$ the type of hypergeometric function appearing in the fundamental period reduces to $_{p-1}F_{q-1}$.}

Another function we will need is the bilateral hypergeometric function
\begin{equation}
    \pFq{m}{\mathcal{H}}{m}{a_{1},\dots,a_{m}}{b_{1},\dots,b_{m}}{x;\epsilon} := \frac{\prod\limits_{i=1}^m\Gamma[b_i]}{\prod\limits_{i=1}^m\Gamma[a_i]}\sum\limits_{n=-\infty}^\infty \frac{\prod\limits_{i=1}^m\Gamma[a_i+\epsilon+n]}{\prod\limits_{i=1}^m\Gamma[b_i+\epsilon+n]}x^{n+\epsilon}\;.
\end{equation}
These bilateral hypergeometric functions are combinations of usual hypergeometric functions of argument $x$ and $1/x$. But it turns out that in the $\epsilon$-expansion to order $\mathcal{O}(\epsilon^{m})$ only the hypergeometric function with argument $x$ contributes, as given by the identity \cite{Zudilin:2018joh}
\begin{equation}
\begin{aligned}
 &\pFq{m}{\mathcal{H}}{m}{a_{1},\dots,a_{m}}{1,\dots,1}{x;\epsilon} = \frac{1}{\prod\limits_{i=1}^m\Gamma[a_i]}\sum\limits_{n=-\infty}^\infty \frac{\prod\limits_{i=1}^m\Gamma[a_i+\epsilon+n]}{\prod\limits_{i=1}^m\Gamma[1+\epsilon+n]}x^n=\\
 &=\frac{z^\epsilon\prod\limits_{i=1}^{m}\Gamma[a_i+\epsilon]}{\Gamma[1+n]\prod\limits_{i=1}^{m}\Gamma[a_i]}\;\pFq{m+1}{F}{m}{1,a_{1}+\epsilon,a_{2}+\epsilon,a_{3}+\epsilon}{1+\epsilon,1+\epsilon,1+\epsilon}{x} + \mathcal{O}(\epsilon^{m})\;.
\end{aligned}
\label{eq:bihypepsilon}
\end{equation}
The hypergeometric function on the right hand side of this equation is exactly of the form appearing as a solution of the GKZ system of CY manifolds. Thus, a combination of its first $m-1$ parameter derivatives and the hypergeometric function itself can be rewritten in terms of the $\epsilon$-expansion of a bilateral hypergeometric function to order $\mathcal{O}(\epsilon^{m})$. Interestingly, this correspondence holds exactly to the required order to compute all periods.

The $\epsilon$-expansion of certain bilateral hypergeometric functions takes a very simple form. In this paper we specialize to the parameter values $a_i=\{1/2,a,1-a\}$. We define $F_{i}(x)$ and $H_{i}(x)$ functions as the ones appearing at order $\epsilon^{i}$ in \eqref{eq:bihypepsilon} for $m = 2$ and $m = 3$ respectively, i.e.\
\begin{align}
    F(x;\epsilon) &:= \pFq{2}{\mathcal{H}}{2}{a,1-a}{1,1}{x;\epsilon} = F_{0}(x) + F_{1}(x)\epsilon + \mathcal{O}(\epsilon^{2})\,,\\
    H(x;\epsilon) &:= \pFq{3}{\mathcal{H}}{3}{\frac{1}{2},a,1-a}{1,1}{x;\epsilon} = H_{0}(z) + H_{1}(z)\epsilon + H_{2}(z)\epsilon^{2} + \mathcal{O}(\epsilon^{3})\,.
\end{align}
From the right hand side of \eqref{eq:bihypepsilon} we see that
\begin{align}
    F_{0}(x) &= \pFq{2}{F}{1}{a,1-a}{1}{x}\,,\\
    H_{0}(x) &= \pFq{3}{F}{2}{\frac{1}{2},a,1-a}{1,1}{x}\,.
\end{align}
Then, defining the transformed variable
\begin{equation}
    \tilde{x} := \frac{1}{2} \left( 1-\sqrt{1-x} \right)
\end{equation}
we have Clausen's classical identity
\begin{equation}
    H_{0}(x) = F_{0}(\tilde{x})^{2}\,.
\end{equation}
Thanks to a generalized version of Clausen's identity and to the structure of the bilateral hypergeometric functions involved we also have the relations \cite{Zudilin:2018joh}
\begin{align}
    H_{1}(x) &= F_{0}(\tilde{x})F_{1}(\tilde{x})\,,\\
    H_{2}(x) &= \frac{1}{2} \left( \frac{\pi}{\sin(\pi a)} \right)^{2} F_{0}(\tilde{x})^{2} + F_{1}(\tilde{x})^{2}\,.
\end{align}

We can exploit these identities to express the second parameter derivative in the $\epsilon$-expansion of ${}_{4}{F}_{3}(1,1/2+\epsilon,r+\epsilon,1-r+\epsilon;1+\epsilon,1+\epsilon,1+\epsilon;x)$ in terms of the first parameter derivative and the function itself. Denoting the Euler-Mascheroni constant by $\gamma$ and the polygamma function of order $m$ by
\begin{equation}
    \psi^{(m)} := \frac{d^{m+1}}{dz^{m+1}} \log \left(\Gamma(x)\right)\,,\quad \psi(x) := \psi^{(0)}(x)\,,
\end{equation}
we obtain from the right hand side of \eqref{eq:bihypepsilon}
\begin{align}
    F_{1}(x) &= 2 \gamma F_{0}(x) + F_{0}(x) \log(x) + F_{0}(x) \psi\left(\frac{1}{6}\right) + F_{0}(x) \psi\left(\frac{5}{6}\right)\\
    &+ \left.\frac{d}{d\epsilon}\, \pFq{3}{F}{2}{1,a+\epsilon,1-a+\epsilon}{1+\epsilon,1+\epsilon}{x}\right|_{\epsilon=0}\,,\nonumber\\[2ex]
    H_{1}(x) &= -3 H_{0}(x) \log(12) + H_{0}(x) \log(z)\\
    &+ \left.\frac{d}{d\epsilon}\, \pFq{4}{F}{3}{1,\frac{1}{2}+\epsilon,a+\epsilon,1-a+\epsilon}{1+\epsilon,1+\epsilon,1+\epsilon}{x}\right|_{\epsilon=0}\,,\nonumber\\[2ex]
    H_{2}(x) &= 2 \pi^2 H_{0}(x) + \frac{9}{2} H_{0}(x) \log(12)^2 - 3 H_{0}(x) \log(12) \log(x)\\
    &+ \frac{1}{2} H_{0}(x) \log(x)^2 + \log\left(\frac{x}{1728}\right) \left.\frac{d}{d\epsilon}\, \pFq{4}{F}{3}{1,\frac{1}{2}+\epsilon,a+\epsilon,1-a+\epsilon}{1+\epsilon,1+\epsilon,1+\epsilon}{x}\right|_{\epsilon=0}\nonumber\\
    &+ \left.\frac{1}{2}\frac{d^{2}}{d\epsilon^{2}}\, \pFq{4}{F}{3}{1,\frac{1}{2}+\epsilon,a+\epsilon,1-a+\epsilon}{1+\epsilon,1+\epsilon,1+\epsilon}{x}\right|_{\epsilon=0}\,.\nonumber
\end{align}
Combining all of the above expressions we arrive at
\begin{equation}
\begin{aligned}
    &\left.\frac{d^{2}}{d\epsilon^{2}}\,\pFq{4}{F}{3}{1,\frac{1}{2}+\epsilon,a+\epsilon,1-a+\epsilon,1+\epsilon}{1+\epsilon,1+\epsilon}{x}\right|_{\epsilon=0} =\\
    &= \left( \frac{H_{1}(x)}{\sqrt{H_{0}(x)}} + \sqrt{H_{0}(x)} \log(1728) - \sqrt{H_{0}(x)} \log(x) \right)^{2}\\
    &= \frac{\left(\left.\frac{d}{d\epsilon}\, \pFq{4}{F}{3}{1,\frac{1}{2}+\epsilon,a+\epsilon,1-a+\epsilon}{1+\epsilon,1+\epsilon,1+\epsilon}{x}\right|_{\epsilon=0}\right)^{2}}{H_{0}(x)}\,.
\end{aligned}
\end{equation}

For practical computations the problem remains of obtaining a closed expression for
\begin{equation*}
    \left.\frac{d}{d\epsilon}\, \pFq{4}{F}{3}{1,\frac{1}{2}+\epsilon,a+\epsilon,1-a+\epsilon}{1+\epsilon,1+\epsilon,1+\epsilon}{x}\right|_{\epsilon=0}\,.
\end{equation*}
We can make the observation that the first $\epsilon$-derivative always leaves one of the $b_{i} = 1 + \epsilon$ parameters intact, that after setting $\epsilon = 0$ cancels with $a_{1} = 1$. Therefore, we can express it as
\begin{equation}
    \scalemath{0.96}{
    \left.\frac{d}{d\epsilon}\, \pFq{4}{F}{3}{1,\frac{1}{2}+\epsilon,a+\epsilon,1-a+\epsilon}{1+\epsilon,1+\epsilon,1+\epsilon}{x}\right|_{\epsilon=0} = \left.\frac{d}{d\epsilon}\, \pFq{3}{F}{2}{\frac{1}{2}+\epsilon,a+\epsilon,1-a+\epsilon}{1+2\epsilon,1+\epsilon}{x}\right|_{\epsilon=0}\,,
    }
\vrule width0pt depth20pt
\end{equation}
where $b_{1} = 1+2\epsilon$ accounts for combinatorics of the terms appearing in the derivative of the ${}_4F_{3}$. We then further use the identity
\begin{equation}
    \pFq{3}{F}{2}{a,b,\frac{a+b}{2}}{\frac{a+b+1}{2},a+b}{x} = \pFq{2}{F}{1}{\frac{a}{2},\frac{b}{2}}{\frac{a+b+1}{2}}{x}^{2}
\end{equation}
to obtain
\begin{equation}
    \left.\frac{d}{d\epsilon}\, \pFq{4}{F}{3}{1,\frac{1}{2}+\epsilon,a+\epsilon,1-a+\epsilon}{1+\epsilon,1+\epsilon,1+\epsilon}{x}\right|_{\epsilon=0} = \left.\frac{d}{d\epsilon}\, \left( \pFq{2}{F}{1}{\frac{a+\epsilon}{2},\frac{1-a+\epsilon}{2}}{1+\epsilon}{x} \right)^{2}\right|_{\epsilon={0}}\,,
\vrule width0pt depth25pt
\label{eq:4F3to2F1der}
\end{equation}
thus reducing the problem to the computation of the first parameter derivative of a ${}_2F_{1}$ hypergeometric function.

The derivative with respect to the upper parameters can be found in \cite{2018arXiv180804837B}, where the closed expression
\begin{equation}
    \left.\frac{d}{d\epsilon}\, \pFq{2}{F}{1}{a+\epsilon,b+\epsilon}{a+b}{x}\right|_{\epsilon=0} = \log\left(\frac{1}{1-x}\right) \pFq{2}{F}{1}{a,b}{a+b}{x} =: u(a,b;x)
\end{equation}
is given. For the derivative with respect to the lower parameter we find in \cite{2018arXiv180102428N}
\begin{equation}
\begin{aligned}
    &-\left.\frac{d}{dc}\, \pFq{2}{F}{1}{a,1-a}{c}{x}\right|_{c=1} = \frac{\pi}{2 \sin{\pi a}} \pFq{2}{F}{1}{a,1-a}{1}{1-x} + \frac{1}{2} \left( \psi\left(1-\frac{a}{2}\right)\right.\\
    &\left. + \psi\left(\frac{a+1}{2}\right) - \psi(1) - \psi\left(\frac{1}{2}\right) - \frac{\pi}{\sin{\pi a}} - \log \left(\frac{1-x}{x}\right) \right)\pFq{2}{F}{1}{a,1-a}{1}{x}\\
    & = -l(a;x)\,.
\end{aligned}
\vrule width0pt depth40pt
\end{equation}
By applying the chain rule the closed expression for the first $\epsilon$-derivative of the Gaussian hypergeometric function is given by
\begin{equation}
    \left.\frac{d}{d\epsilon}\, \pFq{2}{F}{1}{a+\epsilon,1-a+\epsilon}{1+\epsilon}{x}\right|_{\epsilon=0} = u(a,1-a;x) + l(a;x) =: v(a;x)\,.
\end{equation}
This is not yet of the form of the derivative appearing in \eqref{eq:4F3to2F1der}. To relate the two we can apply the hypergeometric identity
\begin{equation}
    \pFq{2}{F}{1}{\tilde{a},\tilde{b}}{\frac{\tilde{a}+\tilde{b}+1}{2}}{x} =  \pFq{2}{F}{1}{\frac{\tilde{a}}{2},\frac{\tilde{b}}{2}}{\frac{\tilde{a}+\tilde{b}+1}{2}}{4x(1-x)}\,,
\end{equation}
with $\tilde{a} = a+\epsilon$ and $\tilde{b} = 1-a+\epsilon$ and transforming the argument by $x \rightarrow \tilde{x}$. In this way we obtain the closed form for the derivative
\begin{equation}
    \left.\frac{d}{d\epsilon}\, \pFq{4}{F}{3}{1,\frac{1}{2}+\epsilon,a+\epsilon,1-a+\epsilon}{1+\epsilon,1+\epsilon,1+\epsilon}{x}\right|_{\epsilon=0} = 2\, \pFq{2}{F}{1}{a,1-a}{1}{\tilde{x}} v(a;\tilde{x})\,.
\vrule width0pt depth25pt
\label{eq:4F3derclosedform}
\end{equation}

\section{Symmetric square}

The general second order differential operator quadratic in the variable $x$ can be written as
\begin{equation}
    \mathcal{L} = \theta^2-x(A\theta^2+B\theta+C)+x^2 (D \theta^{2} + E \theta + F)\,.
\end{equation}
Computing its symmetric square yields
\begin{equation}
\begin{aligned}
    \tilde{\mathcal{L}} &= \theta ^3 + x \left(-2 A \theta ^3+\theta  (-B-4 C)-3 B \theta ^2-2 C\right)\\
    &+ x^2 \left(\theta ^3 \left(A^2+2 D\right)+\theta  \left(4 A C+2 B^2+2 E+4 F\right)+\theta ^2 (3 A B+3 E)+4 B C+4 F\right)\\
    & + x^3 \left(\theta  (-A E-4 A F+B D-4 B E-4 C D)+\theta ^2 (-3 A E-3 B D)\right.\\
    &\left.-2 A D \theta ^3-2 A F-4 B F+2 C D-4 C E\right)\\
    & + x^4 (D\theta +E) \left(D \theta ^2+2 E \theta +4 F\right)\,.
\end{aligned}
\end{equation}
For certain choices of parameters $\{A,B,C,D,E,F\}$ the symmetric square factorizes by a polynomial $\alpha x^{2} + \beta x + \gamma$ yielding an operator quadratic in $x$. There are four distinct combinations of parameters for which this occurs, which we list below.

\begin{itemize}
    \item \textbf{Case 1:} The choice of parameters
    \begin{equation}
        A = -\frac{\beta  E}{\alpha }\,,\quad B = -\frac{\beta  E}{2 \alpha }\,,\quad D = E\,,\quad \gamma  = \frac{\alpha }{E}\,,
    \end{equation}
    gives the symmetric square
    \begin{equation}
        \begin{aligned}
        \tilde{\mathcal{L}} &= \theta^3 + x \left(-4 C \theta -2 C+\frac{\beta  E \theta ^3}{\alpha }+\frac{3 \beta  E \theta ^2}{2 \alpha }+\frac{\beta  E \theta }{2 \alpha }\right)\\
        &+x^2 \left(E \theta ^3+3 E \theta ^2+2 E \theta +4 F \theta +4 F\right)\,.
        \end{aligned}
    \end{equation}
    
    \item \textbf{Case 2:} The choice of parameters
    \begin{equation}
        A = -\frac{2 \alpha }{\beta }-\frac{\beta  D}{2 \alpha }\,,\quad B = -\frac{\beta  E}{2 \alpha }\,,\quad C = -\frac{\beta  F}{2\alpha }\,,\quad \gamma  = \frac{\beta ^2}{4 \alpha }\,,
    \end{equation}
    gives the symmetric square
    \begin{equation}
        \begin{aligned}
        \tilde{\mathcal{L}} &= \theta ^3 +x\left(\frac{\beta  D \theta ^3}{\alpha }+\frac{3 \beta  E \theta ^2}{2 \alpha }+\frac{\beta  E \theta }{2 \alpha }+\frac{2 \beta  F\theta }{\alpha }+\frac{\beta  F}{\alpha }\right)\\
        &+ x^2 \left(\frac{\beta ^2 D^2 \theta ^3}{4 \alpha ^2}+\frac{3 \beta ^2 D E \theta ^2}{4 \alpha^2}+\frac{\beta ^2 D F \theta }{\alpha ^2}+\frac{\beta ^2 E^2 \theta }{2 \alpha ^2}+\frac{\beta^2 E F}{\alpha ^2}\right)\,.
        \end{aligned}
    \end{equation}
    
    \item \textbf{Case 3:} The choice of parameters
    \begin{equation}
        \begin{aligned}
        &A = -\frac{2 \beta  E}{\alpha }\,,\quad B = -\frac{\sqrt{E \left(\beta ^2 E-2 \alpha ^2\right)}+\beta  E}{2 \alpha }\,,\\
        &C = \frac{\alpha  F}{\sqrt{E \left(\beta ^2 E-2 \alpha ^2\right)}-\beta  E}\,,\quad D = 2 E\,,\quad \gamma  = \frac{\alpha }{2 E}\,,
        \end{aligned}
    \end{equation}
    gives the symmetric square
    \begin{equation}
        \begin{aligned}
        \tilde{\mathcal{L}} &= \theta ^3 + x \left(\frac{3 \theta ^2 \sqrt{\beta ^2 E^2-2 \alpha ^2 E}}{2 \alpha}+\frac{\theta  \sqrt{\beta ^2 E^2-2 \alpha^2 E}}{2 \alpha }+\frac{2 F \theta  \sqrt{\beta ^2 E^2-2 \alpha ^2 E}}{\alpha  E}\right.\\
        &\left.+\frac{F \sqrt{\beta ^2 E^2-2 \alpha ^2 E}}{\alpha  E}+\frac{2 \beta  E \theta ^3}{\alpha }+\frac{3 \beta  E \theta ^2}{2 \alpha }+\frac{\beta  E \theta }{2 \alpha }+\frac{2 \beta  F \theta }{\alpha }+\frac{\beta  F}{\alpha }\right)\\
        & +x^2 \left(2E\theta^3+3 E \theta ^2+E \theta +4 F \theta +2 F\right)\,.
        \end{aligned}
    \end{equation}
    
    \item \textbf{Case 4:} The choice of parameters
    \begin{equation}
        A = -\frac{\beta  D}{\alpha }\,,\quad B = C = E = F = 0\,,\quad \gamma = \frac{\alpha }{D}\,,
    \end{equation}
    gives the symmetric square
    \begin{equation}
        \tilde{\mathcal{L}} = \theta ^3 + x \left(\frac{\beta  D \theta ^3}{\alpha }\right) + x^2 \left(D \theta ^3\right)\,.
    \end{equation}
\end{itemize}

\clearpage
\bibliography{literature}
\bibliographystyle{utphys}

\end{document}